\newcommand{\nn}{\nonumber}
\newcommand{\nb}{\nabla}

       \newcommand{\Cc}{ {\mathcal{C}} }
       
       \newcommand{\Ec}{ {\mathcal{E}} }
       \newcommand{\Fc}{ {\mathcal{F}} }
       \newcommand{\Gc}{ {\mathcal{G}} }
       \newcommand{\Hc}{ {\mathcal{H}} }
       
       \newcommand{\Jc}{ {\mathcal{J}} }
       \newcommand{\Kc}{ {\mathcal{K}} }
       
       \newcommand{\Mc}{ {\mathcal{M}} }

       \newcommand{\bv}{{\mathbf v}}

       \newcommand{\bB}{{\mathbf B}}

    \newcommand{\hz}{ {\hat{z}} }

    \documentclass[5p,times,twocolumn]{elsarticle}

\usepackage{amssymb}
\usepackage{amsmath}
 \usepackage{amsthm}
 \usepackage{color}

\journal{Physics Letters A}

\begin{document}

\begin{frontmatter}

\title{2D magnetofluid models constructed by a priori imposition of conservation laws}

\author{D. A. Kaltsas}

\ead{dkaltsas@cc.uoi.gr}

\author{G. N. Throumoulopoulos}
 \ead{gthroum@uoi.gr}

\address{%
Department of Physics, University of Ioannina\\
GR 451 10 Ioannina, Greece
}%

\date{\today}
\begin{abstract}
Motivated by a geometric method employed for the derivation of the Nambu bracket for ideal two-dimensional incompressible hydrodynamics, we reconstruct the reduced magnetohydrodynamic (RMHD) model by a priori imposition of its conservation laws. It turns out that there exists a more general model that conserves the same quantities with RMHD. The noncanonical Hamiltonian and Nambu description of this generic system are derived and in addition a canonical description is formed by Clebsch-parameterizing the vorticity and the magnetic flux function. The method for the construction of the dynamical equations is based on the imposition of the conservation laws as orthogonality conditions. Furthermore, this approach enabled us to construct three families of models that respect any combination of two out of the three conservation laws. Some of these models can serve as conservative regularizations of RMHD since under certain conditions they keep the enstrophy bounded without the need of introducing viscosity and may be candidates for incorporating small length scale physics into the RMHD framework.

\end{abstract}



\end{frontmatter}
\section{Introduction}
\label{sec_I}
 The Hamiltonian formulation of fluid models \cite{morr_revi}, such as MHD \cite{morr_gree}, within the Eulerian viewpoint is expressed in terms of noncanonical Poisson brackets. The degeneracy of those brackets gives rise to functionals that Poisson-commute with any arbitrary functional defined on the phase space, the so-called Casimirs, which are invariants that express additional conservation laws. The Casimirs introduce dynamical constraints and together with the Hamiltonian determine the manifold on which the evolution of the dynamical system is restricted.
In \cite{blen_badi} and \cite{blen_badi_2} the authors constructed Nambu-like brackets (trilinear antisymmetric brackets) for the 2D ideal incompressible hydrodynamics and for the shallow water equations respectively, using differential 2-forms to impose orthogonality conditions arising from the conservation of the energy and the Casimirs of the respective models. Here we adopt an analogous approach in the context of the Reduced MHD (RMHD) model. In addition we ascertain that this procedure, i.e. the imposition of the conservation laws as orthogonality constraints, can be useful to derive models possessing Nambu-like and Poisson structures with the a priori definition of two ingredients: 1) the dynamical variables, 2) the functional quantities that are to be conserved by the dynamics. This also provides the freedom to select a subset of the original orthogonality conditions, so as to obtain models that do not conserve all of the ideal invariants, though they are non-dissipative.  Such an idea could potentially be linked with the concept of selective decay in magnetohydrodynamic turbulence, which assumes that the total energy is minimized, subject to the conservation of the helicities \cite{mathaus},  in the sense that one can incorporate additional contributions that break the conservation of a particular invariant in order to regulate its decay individually without affecting the rest of the invariants as the diffusive terms do.
Also, as it is pointed out in section \ref{sec_IV} some classes of those models can serve as useful regularizations of the original RMHD model in the sense that they prevent vorticity singularities while preserving their non-dissipative structure.  Similar regularizations were introduced in \cite{thya} for the three dimensional incompressible MHD equations. 

It is important to emphasize that the present contribution does not completely solve the inverse problem of conservation laws for models of the MHD form. The complete determination of the full set of models that respect given conservation laws would be a very tough task and so far we are not aware of any suitable methodology in order to address the complete problem. As we will see below, we shortcut by considering that the time derivatives of the dynamical variables assume a specific form  compatible with the Lie-Poisson brackets of Hamiltonian systems such as RMHD \cite{morr_haze}. This assumption simplifies significantly the subsequent manipulations.

Beyond the construction of the 2D models, we show how one can derive Nambu representations that describe the dynamics in terms of trilinear brackets. The Nambu formalism for the 3D incompressible MHD has been derived in \cite{sala_kurg} but not for the RMHD model. We should stress here that the kind of Nambu brackets discussed in this paper and also in \cite{sala_kurg} are different from the finite dimensional brackets introduced in the classic paper of Nambu \cite{Nambu1973} since the former are infinite-dimensional analogues of the latter and in general they do not satisfy the generalization of the Jacobi identity for Nambu brackets \cite{Takhtajan1994}. However this kind of infinite-dimensional Nambu-like brackets, introduced initially in \cite{blen_nevi}, are completely antisymmetric and describe dynamics in terms of the Hamiltonian and a second Hamiltonian-like functional which is a Casimir invariant. This provided, in the context of 2D hydrodynamics, a useful tool for constructing preserving algorithms \cite{salmon}, since the Hamiltonian and the Casimir invariant are conserved exactly up to machine precision if one makes sure that the discretization procedure retains the antisymmetry property. Hence, it is of interest to derive the respective RMHD Nambu brackets as a step towards the construction of analogous Casimir preserving algorithms for plasma dynamics.

The plan of the paper is as follows: in section \ref{sec_II} we review the RMHD model, its Hamiltonian formalism and we present its Casimir invariants. In section \ref{sec_III} we impose the RMHD conservation laws (CLs) as orthogonality conditions between the time evolution vector of the dynamical variables and the phase space gradients of the invariants. Within this section the Poisson and Nambu brackets and a canonical representation of a generic model that includes RMHD as a special case are derived.  Section \ref{sec_IV} is devoted in the construction of new 2D models that conserve any two out of the three CLs of RMHD. We conclude and summarize the results of this paper in section \ref{sec_V}.
 
\section{The Reduced MHD model}
\label{sec_II}
Reduced MHD models are used to displace the usual 3D MHD equations when a strong guiding magnetic field $\bB_0$ is present, because they are much more simpler in form and thus can be handled more conveniently. The Reduced MHD model can be rigorously derived by performing asymptotic expansion of the MHD equations with the ordering $L_\perp/L_\parallel\sim B_\perp/B_0\sim v/v_A\sim \epsilon$ and $\epsilon$ being a small parameter \cite{stra}. Here $L_\perp$ and $L_\parallel$ are the characteristic length scales perpendicular and parallel to the guiding field respectively, $B_\perp$ and $B_0$ are the corresponding magnetic field magnitudes, $v$ is the magnitude of the velocity and $v_A$ is the Alfv\'en velocity. Alternatively one can just confine the dynamics to take place on the plane perpendicular to the guiding field $\mathbf{B}_0=B_0 \hz$, i.e. to express $\bB$ and $\bv$ as 
\begin{eqnarray}
\bB=\nb\psi\times\hz\,,\qquad \bv=\nb\chi\times \hz\,, \label{2d_rep}
\end{eqnarray}
where $\psi$ is the magnetic flux function and $\chi$ is the velocity stream function. Assuming that the plasma is incompressible, in the sense that the mass density is uniform throughout the plasma volume, one can derive from the general momentum and induction equations of the MHD model, the following dynamical equations, defined on a bounded domain $D\subset \mathbb{R}^2$,
\begin{eqnarray}
\partial_t \omega&=&[\chi,\omega]+[J,\psi] \,, \label{vort_eq}\\
\partial_t \psi&=&[\chi,\psi]\,, \label{indu_eq}
\end{eqnarray}
where $J:=-\Delta\psi$ and $\omega:=-\Delta \chi$, are the magnitudes of the current density and vorticity respectively and $[a,b]:=(\partial_x a)\,(\partial_y b)-(\partial_x b)\,(\partial_y a)$ is the Jacobi-Poisson bracket. 

In \cite{morr_haze} the authors proved that the RMHD model above, and also its compressible counterpart, possess a noncanonical Hamiltonian structure, since the dynamics can be expressed in terms of a degenerate Poisson bracket and a Hamiltonian, as follows
\begin{eqnarray}
\partial_t\omega=\{\omega,\Hc\}\,, \qquad \partial_t\psi=\{\psi,\Hc\}\,, \label{ham_dyn_eqs}
\end{eqnarray}
where the Hamiltonian $\Hc$ is 
\begin{eqnarray}
\Hc[\omega,\psi]=\frac{1}{2}\int_Dd^2x\, \left(|\nb\chi|^2+|\nb\psi|^2\right)\nn\\
=-\frac{1}{2}\int_D d^2x\, (\omega\Delta^{-1}\omega+\psi \Delta\psi)\,, \label{hamiltonian}
\end{eqnarray}
and the Poisson bracket is given by
\begin{eqnarray}
\hspace{-0.7cm}\{F,G\}=\int_D d^2x\,\left\{\omega[F_\omega,G_\omega]+\psi\left([F_\psi,G_\omega]-[G_\psi,F_\omega]\right)\right\}\,. \label{pois_rmhd}
\end{eqnarray}
Here $F_u:=\delta F/\delta u$ represents the functional derivative of $F$ with respect to the variable $u$. The bracket \eqref{pois_rmhd} has two families of Poisson-commuted functionals, i.e. Casimir invariants, given by
\begin{eqnarray}
\Cc=\int_Dd^2x\, \omega \Fc(\psi)\,,\quad \Mc= \int_Dd^2x\, \Gc(\psi) \,, \label{cas_rmhd}
\end{eqnarray}
where $\Fc$ and $\Gc$ are arbitrary functions. The Casimir $\Cc$ is a cross-helicity-like functional and $\Mc$ expresses the conservation of the magnetic flux.


Therefore the incompressible RMHD model has three general CLs, expressed through the preservation of the functionals $\Hc$, $\Cc$ and $\Mc$. Here, the structure of the dynamical equations \eqref{vort_eq}-\eqref{indu_eq} and of the Poisson bracket \eqref{pois_rmhd} indicated the conservation laws. In this work we try to reverse this procedure, i.e. with the CLs at hand, we try to construct the dynamical equations that conserve the associated invariants.

\section{Construction of the RMHD model from given CLs}
\label{sec_III}
\subsection{Dynamics via orthogonality constraints}
Let us assume that we have a continuous system bounded in a 2D domain $D$ and described by the dynamical variables $X=(X_1,...,X_N)$ that exhibits conservation of a set of M quantities  $Y_1[X],...,Y_M[X]$ expressed as functionals defined on phase space. The conservation of $Y_i[X]$, $i=1,...,M$ yields
\begin{eqnarray}
c_1: \,\frac{d Y_1[X]}{dt}&=&\int_D d^2x\, \frac{\delta Y_1}{\delta X_i}\partial_tX_i=0 \,, \nn\\
&\vdots& \nn \\
c_M:\,\frac{dY_M[X]}{dt}&=&\int_D d^2x\, \frac{\delta Y_M}{\delta X_i}\partial_tX_i=0 \,, \label{orth_cond}
\end{eqnarray}
The equations above define a set of $M$ orthogonality conditions $c_i$, $i=1,...,M$, of the vectors
\begin{eqnarray}
\mu_i=\left(\frac{\delta Y_i}{\delta X_1},...,\frac{\delta Y_i}{\delta X_N}\right)\,, \quad i=1,...,M \,,
\end{eqnarray}
with the vector 
\begin{eqnarray}
\sigma=\left(\partial_t X_1,...,\partial_t X_N \right)^T\,.
\end{eqnarray}
Since $\mu_i$ are known, then in principle the orthogonality conditions \eqref{orth_cond} can be exploited in order to find the components of $\sigma$. In the case of 2D hydrodynamics the orthogonality conditions correspond to the conservation of kinetic energy $\Kc$ and enstrophy $\Ec$. In \cite{blen_badi} the authors imposed conveniently those conditions using differential 2-forms identifying $\Kc_\omega$ as a 0-form $\mu_1$ and $\Ec_\omega$ as a 0-form $\mu_2$. Then $\partial_t\omega=d\mu_1\wedge d\mu_2$, if $d\mu_1\wedge d\mu_2$ is assumed to be exact, is automatically orthogonal to $\mu_1$ and $\mu_2$.

Inspired by the reference \cite{blen_badi}, we adopt a similar approach, i.e. we start by considering the conservation laws as orthogonality conditions like those in \eqref{orth_cond} that act as constraints on the dynamics, in order to construct a 2D continuum model, with dynamical variables the vorticity and the magnetic flux function, that conserves $\Hc$, $\Cc$ and $\Mc$. To do so let us define the following vectors
\begin{eqnarray}
 \xi&:=&\left(\omega,\,\psi\right)\,,\nn \\
 f&:=&\left(\Hc_\omega,\,\Hc_\psi\right)=\left(\chi,\,J\right)\,,\nn\\
 g&:=&\left(\Cc_\omega,\,\Cc_\psi\right)=\left(\Fc(\psi),\,\omega \Fc'(\psi)\right)\,,  \nn \\
 h&:=&\left(\Mc_\omega,\,\Mc_\psi\right)=\left(0,\,\Gc'(\psi) \right)\,,\nn \\
 \sigma&:=&\partial_t\xi^T=\left(\partial_t\omega,\,\partial_t\psi\right)^T, \label{vecs}
\end{eqnarray}
The time invariance of $\Hc$, $\Cc$ and $\Mc$ yields
\begin{eqnarray}
c_1&:&\, \frac{d\Hc}{dt}=\int_D d^2x\, \left[\Hc_\omega(\partial_t\omega)+\Hc_\psi (\partial_t\psi) \right]=0\,,\nn\\
c_2&:&\, \frac{d\Cc}{dt}=\int_D d^2x\, \left[\Cc_\omega(\partial_t\omega)+\Cc_\psi (\partial_t\psi) \right]=0\,,\nn\\
c_3&:&\, \frac{d\Mc}{dt}= \int_D d^2x\,\left[ \Mc_\omega(\partial_t\omega) +\Mc_\psi (\partial_t\psi)\right]=0\,,
\end{eqnarray}
Those orthogonality conditions can be expressed with the use of Eqs. \eqref{vecs} as 
\begin{eqnarray}
c_1&:&\,\, \int_D d^2x\, f_i \sigma_i=0\,, \nn \\
 c_2&:&\, \, \int_D d^2x\, g_i \sigma_i=0\,,\nn\\
  c_3&:&\, \, \int_D d^2x\, h_i\sigma_i=0\,.
\end{eqnarray}
In noncanonical Hamiltonian theories involving Lie-Poisson brackets the dynamics is governed by Hamilton's equations of the form 
\begin{eqnarray}
\partial_t F=\{F,\Hc\}\,, \label{hamilton_eqs}
\end{eqnarray}
with $\{F,G\}$ being a Lie-Poisson bracket with  general form 
\begin{equation}
\{F,G\}=\int_Dd^2x\, \xi^k [F_{\xi^m},G_{\xi^n}]W_k^{m,n}\,, \label{gen_lie_poisson}
\end{equation} 
where $\xi$ is the set of the dynamical variables and $W_k^{m,n}$ are constants. 
Equations \eqref{hamilton_eqs}, \eqref{gen_lie_poisson} indicate that the time independent parts of the evolution equations can be written as linear combinations of Jacobi-Poisson brackets between the various dynamical variables and the functional derivatives of the Hamiltonian. Therefore we assume that $\sigma_{1,2}$ can be expanded as 
\begin{eqnarray}
\sigma_i=\gamma_{ijk}\,[f_j,\xi_k]\,,\quad i,j,k=1,2\,. \label{2form}
\end{eqnarray}
This Ansatz for the quantities $\sigma_i$ possibly excludes models that do respect the given CLs but the time derivatives of the dynamical variables do not assume this specific form. However it  is consistent within the Lie-Poisson Hamiltonian framework and is a helpful assumption in order to carry out manipulations that easily result in systems of equations that preserve the Casimirs. In fact exploiting the identity
\begin{eqnarray}
\int_D d^2x\, a[b,c]=\int_D d^2x\, c[a,b]=\int_D d^2x\, b[c,a]\,, \label{identity}
\end{eqnarray}
which holds for appropriate boundary condition e.g. periodic, we can find that the constraints $c_1$, $c_2$ and $c_3$ induce the following sets of conditions for the parameters $\gamma_{ijk}$ with $i,j,k=1,2$, which have to hold true for the dynamical variables $\psi$ and $\omega$ to be independent, 
\begin{eqnarray}
c_1&:&\{\gamma_{121}=\gamma_{211}\,, \; \gamma_{122}=\gamma_{212}\}\,, \nn \\
c_2&:& 
\{\gamma_{111}=\gamma_{212}\,,\; \gamma_{121}=\gamma_{222}\}\,,\;  \Fc''(\psi)=0\,, \nn \\ 
&&  \{\gamma_{111}=\gamma_{212}\,,\; \gamma_{121}=\gamma_{222}\,,\nn \\
&&  \gamma_{211}=0\,,\; \gamma_{221}=0\}\,, \; \Fc''(\psi)\neq 0\,, \nn\\
c_3&:&\{\gamma_{211}=0\,,\; \gamma_{221}=0\}\,, \label{parameter_sets}
\end{eqnarray} 
and the rest parameters $\gamma_{ijk}$ are arbitrary. The constraint $c_3$, stemming from the conservation of the magnetic flux, contributes in determining the parametric conditions only when $\Fc(\psi)$ is linear in $\psi$. In all other cases, imposing the conservation of the cross helicity functional ensures the conservation of the magnetic flux as well. Therefore the trajectory of the RMHD dynamics is determined by the intersection of the energy and the cross helicity level sets in phase space if $\Fc''(\psi)\neq 0$.  

The above results imply that the conservation of the RMHD Casimirs require
\begin{eqnarray}
\gamma_{111}=\gamma_{212}=\gamma_{122}\equiv \epsilon_1\,,\quad \gamma_{112}\equiv\epsilon_2\,,\nn\\
\quad \gamma_{121}=\gamma_{211}=\gamma_{221}=\gamma_{222}=0, \label{book1}
\end{eqnarray} 
where $\epsilon_1$ and $\epsilon_2$ are arbitrary parameters. We introduce these new parameters in order to simplify the notation. In view of \eqref{book1} and \eqref{2form} we take
\begin{eqnarray}
\partial_t\omega&=& \left([\Hc_\psi,\psi]+[\Hc_\omega,\omega]\right)+\epsilon[\Hc_\omega,\psi]\,, \label{dyn_eq_fun_der_1}\\
\partial_t\psi&=& [\Hc_\omega,\psi]\,. \label{dyn_eq_fun_der_2}
\end{eqnarray}
where the parameter $\epsilon_1$ was absorbed by rescaling the time variable, and $\epsilon_2$ was renamed. Evaluating the functional derivatives we obtain
\begin{eqnarray}
\partial_t\omega&=&([J,\psi]+[\chi,\omega])+\epsilon[\chi,\psi]\,, \label{inclusive_1}\\
\partial_t\psi&=&[\chi,\psi]\,.\label{inclusive_2}
\end{eqnarray}
It is easy to corroborate that the model \eqref{inclusive_1}-\eqref{inclusive_2} conserves the energy $\Hc$ and the Casimirs $\Cc$ and $\Mc$ as given in \eqref{cas_rmhd}. This generalized model includes the RMHD as a special case since the latter is recovered for $\epsilon=0$. 
Note that the conditions \eqref{book1} can be interpreted as follows: the inclusion of $[J,\omega]$ in any of the dynamical equations of the model violates the conservation laws. In addition, the evolution of $\psi$ is coerced to contain no dependence on $J$ and $\omega$.
\subsection{Poisson and Nambu bracket description}
\label{sub_sec_III_2}
For deriving the Poisson and Nambu brackets for \eqref{inclusive_1}--\eqref{inclusive_2} we have just to consider the time evolution of an arbitrary functional $F=F[\omega,\psi]$ 
\begin{eqnarray}
\partial_t F=\int_D d^2x\, \left[F_\omega (\partial_t \omega)+F_\psi (\partial_t\psi)\right]\,, \label{dtF}
\end{eqnarray}
and use the equations \eqref{dyn_eq_fun_der_1}, \eqref{dyn_eq_fun_der_2}, with the arbitrary functional $G$ replacing the Hamiltonian, to obtain
\begin{eqnarray}
\{F,G\}=\{F,G\}_{RMHD}+\epsilon\int_Dd^2x\,\psi [F_\omega,G_\omega]\,,  \label{pois_I}
\end{eqnarray}
where $\{F,G\}_{RMHD}$ is given by \eqref{pois_rmhd}. Bracket \eqref{pois_I} satisfies the Jacobi identity since the matrices $W^n$, $n=1,2$ in \eqref{gen_lie_poisson} pairwise commute \cite{tif}.
For the derivation of the Nambu formalism of the system \eqref{inclusive_1}--\eqref{inclusive_2} we need just to observe that $\psi=\bar{\Cc}_\omega$ and $\omega=\bar{\Cc}_\psi$ where $\bar{\Cc}=\Cc$ for $\Fc(\psi)=\psi$. Making this substitution we convert the Lie-Poisson bracket \eqref{pois_I} to the following trilinear bracket
\begin{eqnarray}
\{F,G,Z\}:=\int_D d^2x\, \big\{F_\omega [G_\psi,Z_\omega]+ F_\omega [G_\omega,Z_\psi]\nn \\
+ F_\psi[G_\omega,Z_\omega]+\epsilon F_\omega [G_\omega,Z_\omega]\big\}\,. \label{nambu_1}
\end{eqnarray}
The dynamics can be described by $\partial_tF=\{F,\Hc,\bar{\Cc}\}$.  The bracket is completely antisymmetric in its three arguments in view of the identity \eqref{identity} and the antisymmetry of the Jacobi-Poisson bracket $[f,g]=-[g,f]$. The Nambu representation can be proved useful in constructing numerical schemes that preserve up to machine precision the energy and the cross helicity \cite{salmon}. Note though that in our case such a numerical scheme would ensure the conservation of $\bar{\Cc}$ only and not of the entire family $\Cc$ nor the conservation of $\Mc$.
\subsection{Canonical description}
One may derive a canonical description of the system \eqref{inclusive_1}--\eqref{inclusive_2} by expressing the vorticity $\omega$ and the flux function $\psi$ in terms of Clebsch potentials. Canonical descriptions of the RMHD model were derived in \cite{morr_haze} and in \cite{kane_yosh}. The former derivation needs four Clebsch potentials while the latter only two and both assumed the vorticity to be a Clebsch 2-form $\omega=[P,Q]$. In \cite{kane_yosh} the authors found two suitable parametrization schemes for $\psi$ namely $\psi=P^\alpha Q^\beta$ and $\psi=P^\alpha+Q^\beta$. Following \cite{kane_yosh} we use $\psi=P^\alpha Q^\beta$ for our generic model and also we make a necessary modification in parameterizing $\omega$
\begin{eqnarray}
\omega&=&[P,Q]+\epsilon P^\alpha Q^\beta \,, \nn\\
\psi&=& P^\alpha Q^\beta\,. \label{canonization}
\end{eqnarray}
The Hamiltonian takes the form 
\begin{eqnarray}
&&\Hc=-\frac{1}{2}\int d^2x \big\{ [P,Q]\Delta^{-1}[P,Q]\nn\\
&&\hspace{0.5cm}+\epsilon P^\alpha Q^\beta \Delta^{-1}[P,Q]
+\epsilon [P,Q] \Delta^{-1}(P^\alpha Q^\beta)\nn\\
&&\hspace{0.5cm}+\epsilon^2 P^\alpha Q^\beta \Delta^{-1}(P^\alpha Q^\beta)+ P^\alpha Q^\beta \Delta(P^\alpha Q^\beta) \big\}\,. \label{canonical_hamiltonian}
\end{eqnarray}
The canonical Hamilton's equations  are
\begin{eqnarray}
\partial_t \left( \begin{array}{c}
P \\Q
\end{array}\right)= \Jc_c \left( \begin{array}{c}
\Hc_P \\ \Hc_Q
\end{array}\right)\,,\label{hamilton_eq}
\end{eqnarray}
where $\Jc_c$ represents the so-called cosymplectic operator 
\begin{eqnarray}
\Jc_c=\left( \begin{array}{ccc}
0&& I\\
-I && 0
\end{array}\right)\,.
\end{eqnarray}
In view of \eqref{canonical_hamiltonian} the Hamilton's equations \eqref{hamilton_eq} take the form 
\begin{eqnarray}
\partial_t P= [\chi,P]+\beta P^\alpha Q^{\beta-1}J+\epsilon \beta P^\alpha Q^{\beta-1}\chi\,, \nn\\
\partial_t Q= [\chi,Q]-\alpha P^{\alpha-1} Q^{\beta}J-\epsilon \alpha P^{\alpha-1} Q^{\beta}\chi\,, \label{canonical_sys}
\end{eqnarray}
where $J=-\Delta(P^\alpha Q^\beta)$ and $\chi=-\Delta^{-1}[P,Q]-\epsilon \Delta^{-1}(P^\alpha Q^\beta)$. Using Eqs. \eqref{canonization} and \eqref{canonical_sys} and exploiting the Jacobi identity for the regular Jacobi-Poisson bracket, one can recover the original system \eqref{inclusive_1}--\eqref{inclusive_2}.
The cross-helicity for $\Fc(\psi)=\psi$ is given by 
\begin{eqnarray}
\Cc=\int_Dd^2x \left(\epsilon P^{2\alpha}Q^{2\beta}+P^\alpha Q^\beta [P,Q]\right)\,. \label{canonical_cas}
\end{eqnarray}
The first term is a conserved quantity due to the conservation of $\Mc$, therefore the second term which is the RMHD cross helicity is also conserved. Hence the conservation of the helicities is retained also on the canonical level. A reason for writing the system \eqref{inclusive_1}-\eqref{inclusive_2} in terms of Clebsch potentials is to see how the addition of the $\epsilon$ term alternates the form of the Hamiltonian. The Clebsch-parameterized Hamiltonian of our generic model is different from its RMHD counterpart, obtained by setting $\epsilon=0$, albeit when expressed in noncanonical Eulerian variables they are identical. This difference is a consequence of the fact that in canonical description any complexity is removed from the Poisson bracket and is transfered in the Hamiltonian. Note that although the Hamiltonian acquires an explicit dependence on the parameter $\epsilon$, the Casimirs do not contain this parameter, since the first term in \eqref{canonical_cas} can be freely subtracted.

\section{Families of reduced models respecting two out of the three original CLs}
\label{sec_IV}
\subsection{$\Hc,\bar{\Cc}$ conserving models}
 From the conditions \eqref{parameter_sets} it is clear that for $\Fc''(\psi)\neq 0$ the most general model that conserves $\Hc$ and $\Cc$ is the model \eqref{inclusive_1}-\eqref{inclusive_2}. However for linear $\Fc(\psi)$ there are a lot of new possibilities since there are two additional arbitrary parameters. The conditions \eqref{parameter_sets} imply that for a family of 2D hydromagnetic models, with dynamical variables the vorticity $\omega$ and the magnetic flux $\psi$, that conserve only the Energy $\Hc$ and the linear cross-helicity $\bar{\Cc}$, the coefficients in the expansions \eqref{2form} should be
\begin{eqnarray}
\gamma_{111}=\gamma_{122}=\gamma_{212}\equiv \epsilon_1\,,\quad \gamma_{112}\equiv\epsilon_2\,,\nn\\
\gamma_{121}=\gamma_{211}=\gamma_{222}\equiv\epsilon_3\,, \quad \gamma_{221}\equiv\epsilon_4\,.\label{book2}
\end{eqnarray} 
Conditions \eqref{book2} with \eqref{2form} lead to the following expansions
\begin{eqnarray}
\partial_t\omega=+\epsilon_1\left([\Hc_\omega,\omega]+[\Hc_\psi,\psi]\right)\nn \\
\epsilon_2[\Hc_\omega,\psi]+\epsilon_3[\Hc_\psi,\omega]\,,\nn\\
\partial_t\psi=\epsilon_1[\Hc_\omega,\psi]+\epsilon_4[\Hc_\psi,\omega]\nn \\
+\epsilon_3\left([\Hc_\omega,\omega]+[\Hc_\psi,\psi]\right)\,, \label{fun_der_gen_model_1}
\end{eqnarray}  
which result in the generalized model
\begin{eqnarray}
\partial_t\omega=\epsilon_1\left([\chi,\omega]+[J,\psi]\right)+\epsilon_2[\chi,\psi]+\epsilon_3[J,\omega]\,,\nn\\
\partial_t\psi=\epsilon_1[\chi,\psi]+\epsilon_3\left([\chi,\omega]+[J,\psi]\right)+\epsilon_4[J,\omega]\,. \label{gen_model_1}
\end{eqnarray}  
The ordinary RMHD model is recovered for $(\epsilon_1,\epsilon_2,\epsilon_3,\epsilon_4)=(1,0,0,0)$. Setting $\epsilon_1=1$ (or rescaling the time variable so as to absorb $\epsilon_1$) in order to retain the RMHD core, we can build, apart from the model \eqref{gen_model_1}, six extensions of RMHD that conserve the Energy and the linear cross-helicity. 
In the generic case represented by the model \eqref{gen_model_1}, $\Mc$ evolves as 
\begin{eqnarray}
\frac{d\Mc}{dt}=\int_Dd^2x\, (\epsilon_3\chi+\epsilon_4J) \Gc''(\psi)[\omega,\psi]\,,
\end{eqnarray}
while evolution of the rest members of the family of cross helicity invariants \eqref{cas_rmhd} is given by 
\begin{equation}
\frac{d\Cc}{dt}=\int_Dd^2x\,(\epsilon_3\chi +\epsilon_4 J )\omega\Fc''(\psi)[\omega,\psi]\,. \label{dCdt}
\end{equation}
Equation \eqref{dCdt} indicates that the conservation of $\Cc$ is possible either if $\Fc''(\psi)=0$, which is the case we discuss in this subsection, or if $\epsilon_3=\epsilon_4=0$, which results in the system \eqref{inclusive_1}--\eqref{inclusive_2} of the previous section, which is consistent with the conditions \eqref{parameter_sets}. 

 The generalized model \eqref{gen_model_1} can be cast into a Hamiltonian form in terms of the Hamiltonian \eqref{hamiltonian} and a Lie-Poisson bracket with Hamilton's equations stemming from the substitution of \eqref{fun_der_gen_model_1} into \eqref{dtF}. By this procedure we find that the Lie-Poisson bracket is 
\begin{eqnarray}
&&\hspace{-1cm}\{F,G\}=\epsilon_1\{F,G\}_{RMHD}+\int_D d^2x \,\Big\{ \epsilon_2 \psi[F_\omega,G_\omega]+\epsilon_4 \omega[F_\psi,G_\psi]\nn \\
&&\hspace{-0.5cm}+\epsilon_3\omega\left([F_\omega,G_\psi]-[G_\omega,F_\psi]\right)+\epsilon_3\psi[F_\psi,G_\psi]\Big\}\,.\label{poisson_gen_1}
\end{eqnarray}
However the Jacobi identity is satisfied only if $\epsilon_1\epsilon_3=\epsilon_2\epsilon_4$. The Nambu description can be obtained similarly with subsection \ref{sub_sec_III_2}, resulting into a completely antisymmetric three-bracket. Note that the requirement $\epsilon_1\epsilon_3=\epsilon_2\epsilon_4$ implies that there are only three non-trivial Hamiltonian extensions of RMHD that conserve $\Hc$ and $\Cc$. As an example let us consider the model $(1,0,0,\epsilon_4)$
\begin{eqnarray}
\partial_t\omega=[\chi,\omega]+[J,\psi]\,,\nn\\
\partial_t\psi=[\chi,\psi]+\epsilon_4[J,\omega]\,. \label{ex_1_1}
\end{eqnarray}  
One can easily identify that \eqref{ex_1_1}, in addition to $\Hc$ and $\Cc$, conserves also a generalized Enstrophy
\begin{eqnarray}
\tilde{\Ec}=\int_D d^2x\, \left(\psi^2+\epsilon_4 \omega^2\right)\,,
\end{eqnarray}
which is a Casimir of the Poisson bracket \eqref{poisson_gen_1} with $\epsilon_2=\epsilon_3=0$. The conservation of this "Enstrophy" functional implies that, if $\epsilon_4 > 0$, this model converts enstrophy to magnetic flux and vice versa, which means that both  $\int_D d^2x \,\omega^2$ and  $\int_D d^2x\, \psi^2$ are bounded  during the evolution, since the maximum value they can attain is the initial value of the invariant $\Ec$. Therefore the addition of the term $\epsilon_4[J,\omega]$ with positive $\epsilon_4$ in the induction equation regularize the RMHD system at least in preventing possible unbounded behavior of the vorticity. Usually such unbounded behavior is remedied by the inclusion of dissipative terms. However dissipation destroys time reversibility and the various conservation laws.

Before proceeding to the next category of models let us make an additional remark: the Poisson bracket \eqref{poisson_gen_1} with $\epsilon_2=\epsilon_3=0$ and $\epsilon_1=1$, incidentally has the same form with the Poisson bracket of a generalized model with finite electron inertia and ion sound Larmor radius effects in 2D geometry \cite{Grasso1999}. Although the bracket has the same form, the evolution equations and the Hamiltonian in the above referenced model are different from \eqref{ex_1_1} and \eqref{hamiltonian} respectively. 
One can see though that the system \eqref{ex_1_1} can be converted to the model with electron inertial and ion sound Larmor radius effects by performing the following transformation $(\psi\rightarrow \psi^*, \chi\rightarrow \chi^*,\omega\rightarrow \omega, J\rightarrow J)$, where $\psi^*=\psi+d_e^2 J$, $\chi^*=\chi+\rho_s^2 \omega$,  and identifying $\epsilon_4=\rho_s^2d_e^2$. Here $d_e$ is the electron skin depth and $\rho_s$ the ion sound Larmor radius. This transformation changes the stream functions but not the corresponding ``vorticities", which means that the fourth and higher order spatial derivatives (associated with very small length scales) are neglected, and leaves the bracket \eqref{poisson_gen_1} identical in form when written in terms of $\omega$ and $\psi^*$ but changes the Hamiltonian. Note also that a similar bracket has been derived for describing the perpendicular dynamics in a 4-field gyrofluid model in \cite{Waelbroeck2012}.

\subsection{$\Hc,\Mc\,-$ conserving models \label{model2}}
To construct models that conserve $\Hc$ and $\Mc$ we need to employ the conditions $c_1$ and $c_3$. According to \eqref{parameter_sets} the imposition of the aforementioned orthogonality conditions leads to 
\begin{eqnarray}
&&\gamma_{111}\equiv \epsilon_1\,,\quad \gamma_{112}\equiv\epsilon_2\,, \quad \gamma_{122}=\gamma_{212}=\epsilon_3\,, \nn \\
&&\hspace{0.5cm}\gamma_{222}\equiv\epsilon_4\,,\quad\gamma_{211}=\gamma_{221}=\gamma_{121}=0\,,
\end{eqnarray}
that is, the general model that conserves $\Hc$ and $\Mc$ is 
\begin{eqnarray}
\partial_t\omega&=&\epsilon_1[\chi,\omega]+\epsilon_2 [\chi,\psi]+\epsilon_3[J,\psi]\,,\nn\\
\partial_t\psi&=&\epsilon_3 [\chi,\psi]+\epsilon_4[J,\psi]\,. \label{gen_model_2}
\end{eqnarray}  
RMHD is recovered for $(\epsilon_1,\epsilon_2,\epsilon_3,\epsilon_4)=(1,0,1,0)$. For the general form of the equations \eqref{gen_model_2} the evolution of the cross-helicity is given by
\begin{eqnarray}
\frac{d\Cc}{dt}=(\epsilon_1-\epsilon_3)\int_D d^2x\,\Fc(\psi) [\chi,\omega]+\epsilon_4\int_D d^2x\, \Fc(\psi) [\omega,J]\,.
\end{eqnarray}
As for the Hamiltonian description, employing the usual procedure of the previous sections for the model \eqref{gen_model_2} we identify that the dynamics is described by \eqref{ham_dyn_eqs} with the following Poisson bracket 
\begin{eqnarray}
&&\{F,G\}=\int_D d^2x\,\big\{(\epsilon_2\psi+\epsilon_1\omega)[F_\omega,G_\omega]\nn \\
&&+\epsilon_3\psi\left([F_\psi,G_\omega]-[G_\psi,F_\omega]\right)+\epsilon_4\psi [F_\psi,G_\psi]\big\}\label{pois_III}\,. 
\end{eqnarray}
The bracket \eqref{pois_III} is clearly antisymmetric and satisfies the Jacobi identity only if $\epsilon_3^2-\epsilon_1\epsilon_3-\epsilon_2\epsilon_4=0$ with roots $\epsilon_3^\pm=\left(\epsilon_1\pm \sqrt{\epsilon_1^2+4\epsilon_2\epsilon_4}\right)/2$. Under this condition, the model \eqref{gen_model_2} has a Hamiltonian structure, with Hamiltonian functional given by \eqref{hamiltonian} and a Poisson bracket given by \eqref{pois_III} which possess
 except of $\Mc$ an additional Casimir which has the form of a  generalized cross helicity 
\begin{equation}
\tilde{\Cc}=\int_D d^2x\, \omega \left(\psi+\frac{\mu_\pm}{2}\omega\right)\,, \label{gen_cros_hel}
\end{equation}
where $\mu_\pm=\left[\left(\epsilon_1\pm \sqrt{\epsilon_1^2+4\epsilon_2\epsilon_4}\right)/(2\epsilon_4)\right]^{-1}$. Now it is known that the absolute value of the cross helicity has the total energy $\Hc$ as an upper bound
(e.g. see \cite{moffatt}), therefore if $\mu_\pm>0$ then the enstrophy is prevented from exhibiting unbounded growth. 

A trilinear bracket formulation is also possible by recognizing that $\omega=\tilde{\Cc}_\psi$ and $\psi=\tilde{\Cc}_\omega-\mu_\pm \tilde{\Cc}_\psi$.  Substituting $\psi$ and $\omega$ in \eqref{pois_III} by these relations we can find a completely antisymmetric trilinear bracket as in subsection \ref{sub_sec_III_2}. The dynamics is described by means of this bracket along with the Hamiltonian and the Casimir $\tilde{\Cc}$.

\subsection{$\Cc, \Mc\,-$ conserving models}
To abandon the requirement for the energy to be conserved, we impose only the constraints $c_2$ and $c_3$. From conditions \eqref{parameter_sets} we take
\begin{eqnarray}
&&\gamma_{111}=\gamma_{212}\equiv \epsilon_1\,, \quad \gamma_{112}\equiv \epsilon_2\,,\quad \gamma_{122}\equiv\epsilon_3\,,\nn \\
&&\gamma_{121}=\gamma_{222}\equiv\epsilon_4\,, \quad \gamma_{211}=\gamma_{221}=0\,,
\end{eqnarray}
that is we obtain the following generalized model
\begin{eqnarray}
\partial_t\omega&=&\epsilon_1[\chi,\omega]+\epsilon_2 [\chi,\psi]+\epsilon_3[J,\psi]+\epsilon_4 [J,\omega]\,,\nn\\
\partial_t\psi&=&\epsilon_1 [\chi,\psi]+\epsilon_4[J,\psi]\,. \label{gen_model_3}
\end{eqnarray}  
RMHD corresponds to $(\epsilon_1,\epsilon_2,\epsilon_3,\epsilon_4)=(1,0,1,0)$. For the generic model \eqref{gen_model_3} the Energy evolution is given by
\begin{eqnarray}
\frac{d\Hc}{dt}=(\epsilon_3-\epsilon_1)\int_Dd^2x\,\chi [J,\psi]+\epsilon_4\int_D d^2x\, \chi [J,\omega]\,. 
\end{eqnarray}
The system \eqref{gen_model_3} conserves the entire families of $\Cc$ and $\Mc$ as given by \eqref{cas_rmhd}.
As an example of a 2D hydromagnetic model that exhibits a selective preservation of the two Casimirs, consider the RMHD generalization $(1,0,1,\epsilon_4)$, i.e.
\begin{eqnarray}
\partial_t\omega&=&[(\chi+\epsilon_4 J),\omega]+[J,\psi]\,,\nn\\
\partial_t\psi&=&[(\chi+\epsilon_4 J),\psi]\,. \label{ex_2}
\end{eqnarray}  
The only difference of \eqref{ex_2} compared to \eqref{vort_eq}--\eqref{indu_eq} is that the former involves in the advection of the fields $\omega$ and $\psi$ the current density $J$. This means that small scale magnetic structures intervene in the advection of $\omega$ and $\psi$, indicating that such or similar models may have some practical implementations in parameterizing subgrid-scale processes when performing numerical simulations or in general short length scale physics, which cannot be described adequately by the RMHD equations.



\section{Conclusion}
\label{sec_V}

We derived a generic 2D hydromagnetic model that conserves the three ideal RMHD invariants $\Hc$, $\Cc$, $\Mc$, by imposing a priori the RMHD conservation laws as orthogonality conditions. The Lie-Poisson and the Nambu brackets, for this generic model follows as simple consequences of the preceding construction procedure. We introduced also a canonized system for this generic model, which can be reduced to a known canonicalization for RMHD.
In addition we found three families of hydromagnetic models that conserve any two out of the three $\Hc$, $\Cc$, $\Mc$.  Some of these or similar models could be candidates for incorporating small length scale physics into the RMHD framework and as conservative regularizations of the RMHD system preventing the flow from forming vorticity singularities without the introduction of viscous terms, which break the conservation properties of the model. Also certain of those models could potentially be useful in regulating the ruggedness of the helicities and the energy individually by introducing small non-dissipative terms associated with small length scale contributions. It is to be proved if the proposed approach to the inverse problem of conservation laws can potentially have other applications e.g. in constructing models from observations, and if some of the models presented above are indeed of physical relevance or practical importance. This could be elaborated by recognizing if the additional contributions can be associated with some parametrization of small length scale physics into the framework of MHD and by performing numerical simulations that could potentially reveal their practical implications.


\section*{Acknowledgements}
The authors would like to thank Dr. Gualtiero Badin for interesting suggestions and for inviting one of the authors (DAK) to the workshop ``Geometric Methods for Geophysical Fluid Dynamics and Climate Modeling'' (Hamburg-Germany, 2017) where part of this work was presented, and also the anonymous reviewers, particularly the second one, for constructive comments which helped to improve substantially the manuscript. DAK was financially supported from the General Secretariat for Research and Technology (GSRT) and the Hellenic Foundation for Research and Innovation (HFRI) (scholarship code: 1817).

\end{document}